\documentclass[fleqn,twoside]{article}
\usepackage{espcrc2}

\usepackage{graphicx}
\usepackage[figuresright]{rotating}

\newcommand{\AmS}{{\protect\the\textfont2
  A\kern-.1667em\lower.5ex\hbox{M}\kern-.125emS}}

\newcommand{\pmt}    {{\sc PMT}}
\newcommand{\mage}    {{\sc MaGe}}
\newcommand{\mpik}    {{\sc MPIK}}
\newcommand{\gerda}    {{\sc Gerda}}
\newcommand{\majorana}    {{\sc Majorana}}

\newcommand{\co}      {{$^{60}$Co}}
\newcommand{\cs}      {{$^{137}$Cs}}
\newcommand{\thorium}      {{$^{232}$Th}}
\newcommand{\thallium}      {{$^{208}$Tl}}
\newcommand{\ra}      {{$^{226}$Ra}}

\newcommand{\onubb}      {{0$\nu\beta\beta$}}

\title{LArGe: Background suppression using liquid argon (LAr) scintillation for 0$\nu\beta\beta$ decay search with enriched germanium (Ge) detectors}

\author{M. Di Marco\address{SNO Institute, Queen's University,
        Kingston (Ontario), K7L3N6, Canada}$^{,b}$,
        P. Peiffer \address[MCSD]{Max-Planck-Institut fuer Kernphysik Heidelberg,
Postfach 10 39 80, 69029 Heidelberg, Germany},
        S.Sch\"onert\addressmark[MCSD]}
       
\begin{document}

\begin{abstract}
Measurements with a bare p-type high purity germanium diode
(HPGe) submerged in a 19~kg liquid argon (LAr) scintillation detector
at \mpik-Heidelberg are reported.  The liquid argon--germanium system
(LArGe) is operated as a 4$\pi$ anti-Compton spectrometer to suppress
backgrounds in the HPGe. This R\&D is carried out in the framework of
the \gerda\ experiment which searches for \onubb\ decays with HPGe
detectors enriched in $^{76}$Ge.  The goal of this work is to develop
a novel method
%, complementary to detector segmentation and 
% pulse shape discrimination, 
to discriminate backgrounds in \onubb\ search which
would ultimately allow to investigate the effective neutrino mass
free of background events down to the inverse mass hierarchy
scale.  Other applications in low-background counting are expected.

%\vspace{1pc}
\end{abstract}

% typeset front matter (including abstract)
\maketitle

\section{Introduction}

The goal of the Germanium Detector Array ({\sc Gerda}) \cite{gerda} is
to search for neutrinoless double beta decays of $^{76}$Ge.  Bare
germanium detectors (HPGe), isotopic enriched in germanium
$^{76}$Ge, will be operated in liquid argon (LAr).  The cryogenic fluid
serves simultaneously as a cooling medium and as a shield against
external radiation. R\&D is carried out to use the scintillation light
of LAr to tag and discriminate backgrounds. The concept and
the proof of principle were first reported in \cite{nu2004}.  The
signature for $0\nu\beta\beta$ decay of $^{76}$Ge is a point-like
energy deposition with $Q_{\beta\beta}=2.039~$MeV inside a HPGe diode.
Background events come mainly from radioactive decays 
and muon induced interactions. These events deposit
typically only a part of their energy inside a HPGe crystal while the residual
energy is dissipated in the adjacent shielding material. Detecting 
the scintillation light of LAr would allow to discriminate these
events.  The work presented here is an R\&D project
within the framework of the {\sc Gerda} experiment.

\section{Experimental setup}

\begin{figure}[htb]
\hspace{9pt}
\includegraphics[height=12pc,width=7pc]{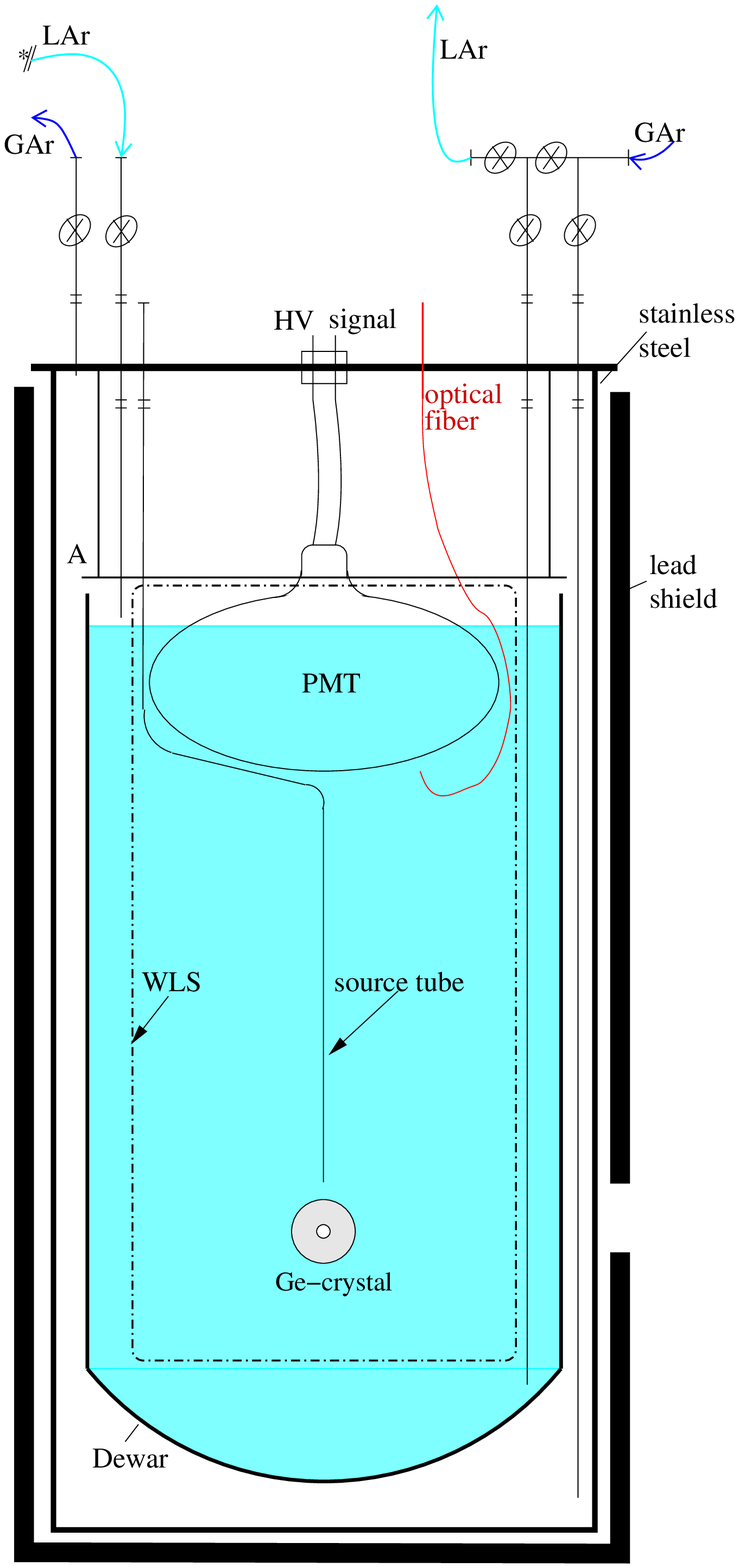}
\includegraphics[width=8pc]{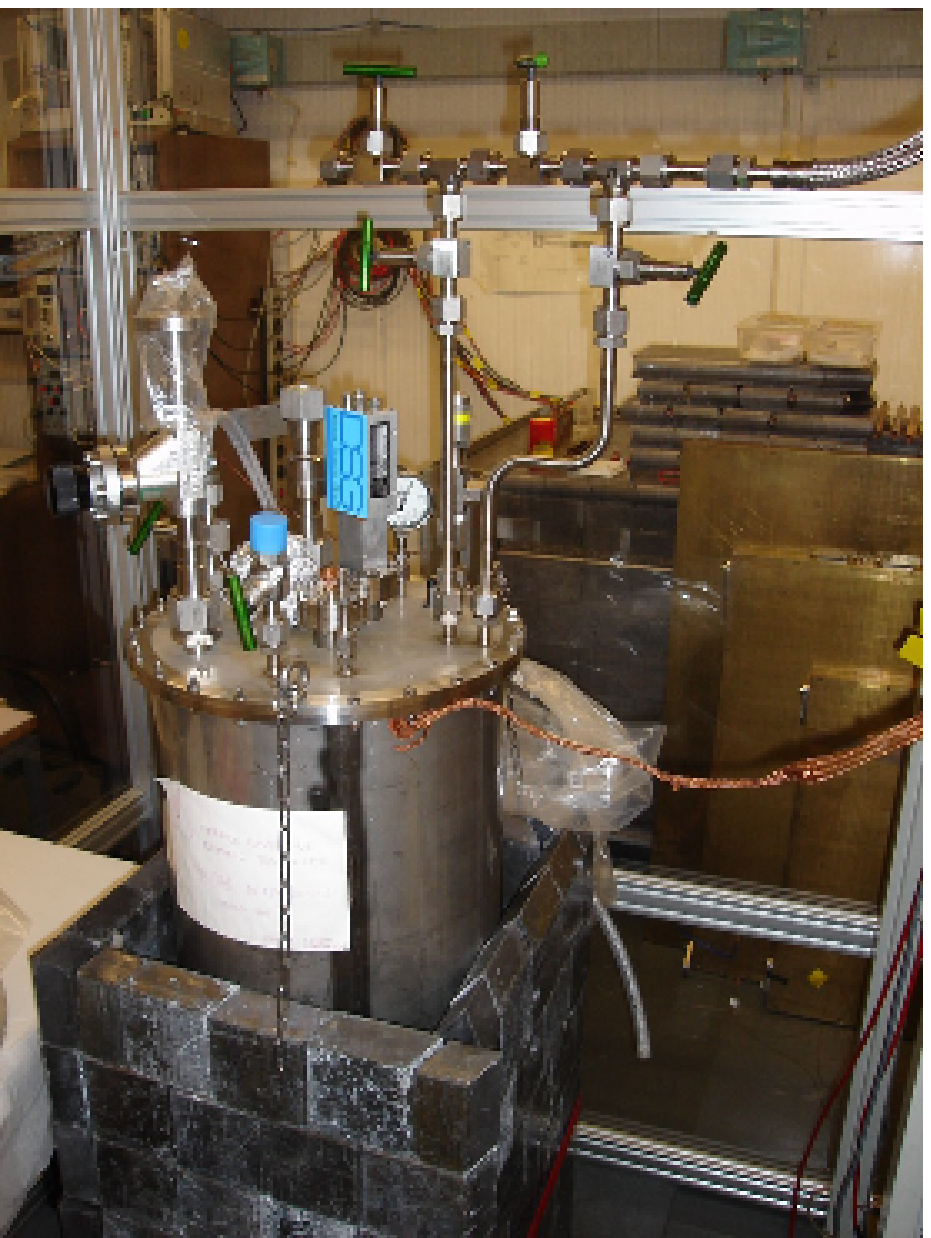}
\caption{Schematic drawing and photo of the LArGe-\mpik~setup.}
\label{fig:setup}
\end{figure}

The experimental setup used for the measurements is displayed in
Fig.~\ref{fig:setup}.  It is located in the
underground laboratory  of the MPIK Heidelberg with an overburden of 
15~mwe.  A bare HPGe
crystal (5.1~cm diameter, 3.5~cm height, 0.39~kg) is submerged in a
dewar (29~cm diameter, 90~cm height) filled with LAr. A wavelength
shifting and reflecting foil defines the active volume of 19~kg LAr. 
The shifted scintillation photons are detected with one 8'' ETL
9357 PMT immersed in LAr.  The dewar system is enclosed in a
gas tight stainless steel tank to prevent quenching from oxygen or
water traces.  Low activity calibration sources can be inserted up to
8~mm from the HPGe crystal via a hermetically closed stainless steel
tube.  The DAQ is triggered by the HPGe diode.  The HPGe and
\pmt~signals are then recorded event-by-event and stored for the
off-line analysis on disk.  HPGe signals are discarded in the analysis
if a simultaneous scintillation signal has been recorded. An analysis
threshold at the single photo electron level was applied. A photo
electron (pe) yield of about 410 pe/MeV was observed during these
measurements.

%%%%%%%%%%%%%%%%%%%%%%%%%%%%%

\section{Measurements}
\begin{figure}[ht]
\includegraphics[width=20pc]{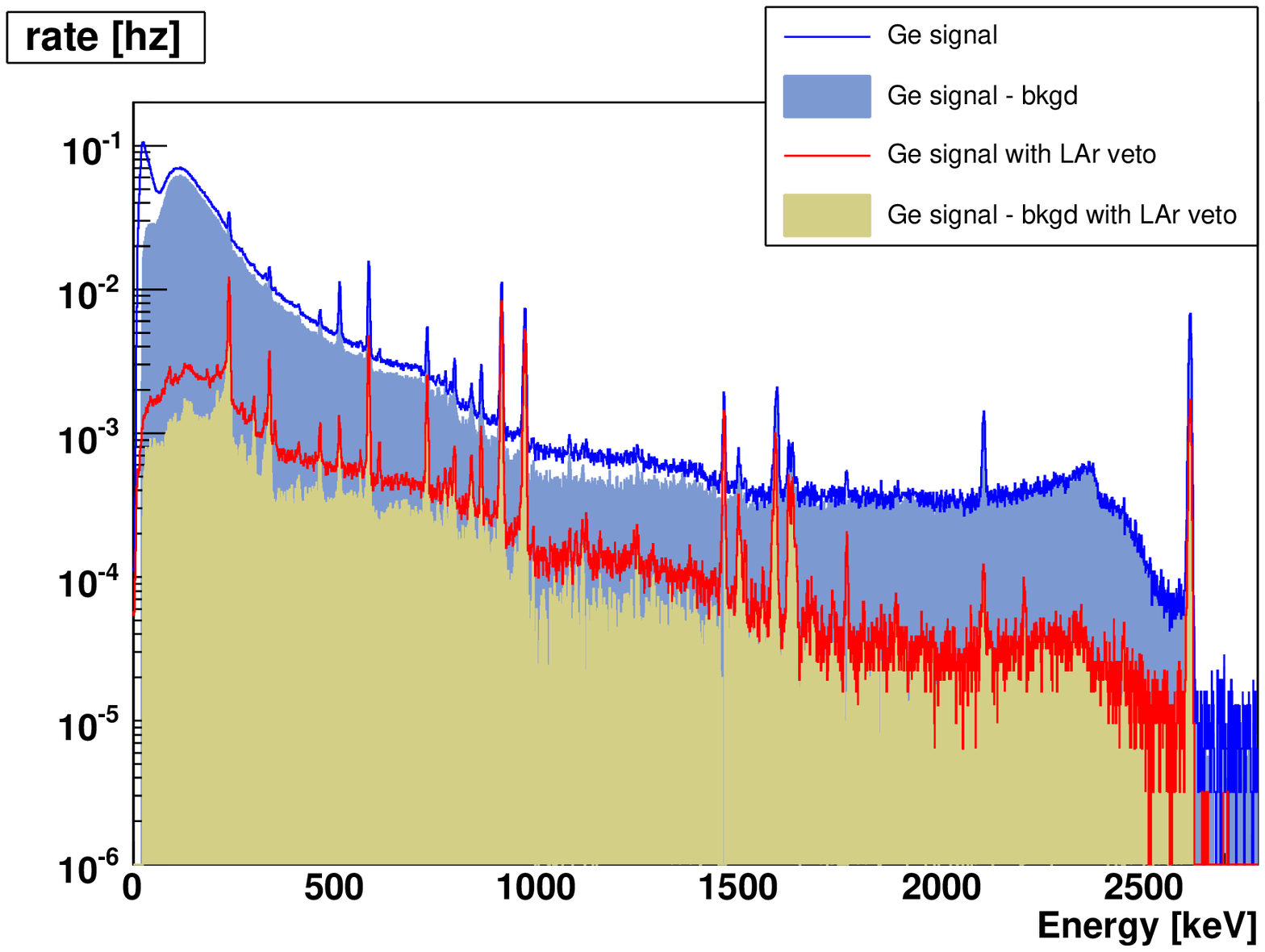}
\includegraphics[width=20pc]{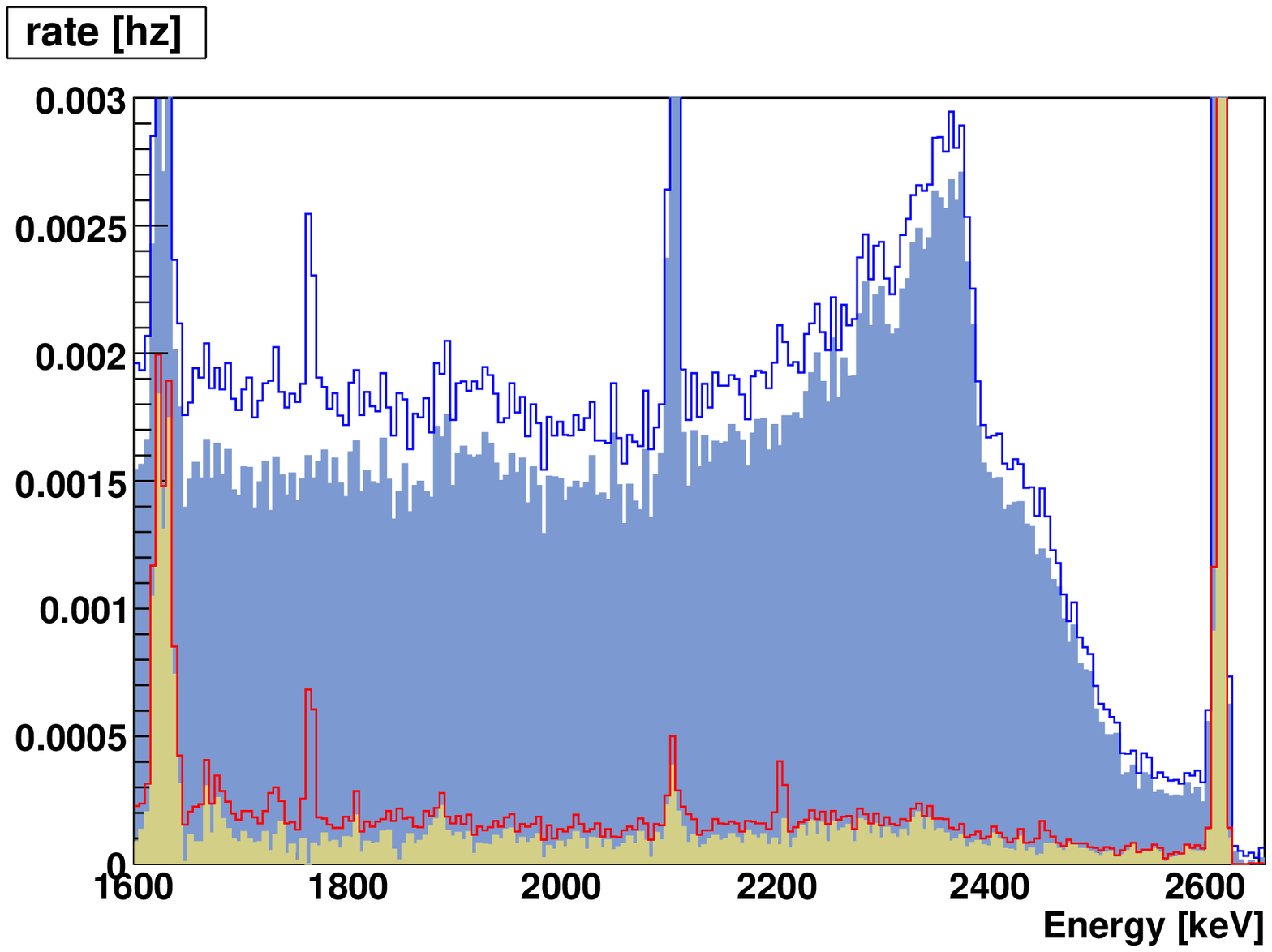}
\caption{Measured \thorium\ energy spectrum 
with (red) and without (blue) LAr anti-coincidence (c.f. text).}
\label{fig:prop}
\end{figure}

The measurement were performed from October to December 2005 using
various gamma sources (\cs, \co, \thorium, \ra), alternated with
periods of background measurements. Given the limited space available in 
this proceedings,
we present only the results achieved with the \thorium\ gamma
source. The source consists of a natural thorium metal wire thus
containing $^{228}$Th and its progeny \thallium . The measured energy
spectrum is displayed in Fig.~\ref{fig:prop}. The line histograms
correspond to spectra without background subtraction and the filled
histograms after background subtraction.  The bottom plot shows a zoom
in the region of interest for \onubb . 
%The LAr veto suppresses about
%94\%  of the Compton continuum at 2039~keV.

The survival probability $P_s$ is defined as the number of counts in a
given energy region after applying the LAr anti-coincidence
cut divided by the number of counts in the same region in the
non-vetoed spectrum and the suppression factor is defined as $S_f = 1-P_s$. 
The background spectra have been measured
separately and subtracted prior to forming the ratio.  For single
gamma decays as for example for the 662~keV \cs\ line, the full energy
(FE) peak is not suppressed after applying the LAr
anti-coincidence cut, since no energy deposition occurs in the LAr. 
The measured value for \cs\ is $P_s = 1.00 \pm 0.01$.  

An important background source for \onubb\ is the photons emitted in
the decay of \thallium , a progeny of \thorium . As the \thallium\
2615~keV gamma is part of a gamma cascade in the deexcitation of \thallium ,
the FE peak is suppressed in case that a second gamma
deposits energy inside the LAr.  $S_f $ for the
2615~keV line is $0.78 \pm 0.01$. \thallium\ Compton events
which deposit an energy close to $Q_{\beta\beta} = 2039$~keV inside
the crystal are vetoed with $S_f=0.94\pm
0.01$, or in other words,  the Compton continuum is suppressed 
by a factor 17.

%%%%%%%%%%%%%%%%%%%%%%%%%%%%%
\section{Simulations}
%%%%%%%%%%%%%%%%%%%%%%%%%%%%%

\begin{figure}[h]
\hspace{9pt}
\includegraphics[angle=90,width=15pc]{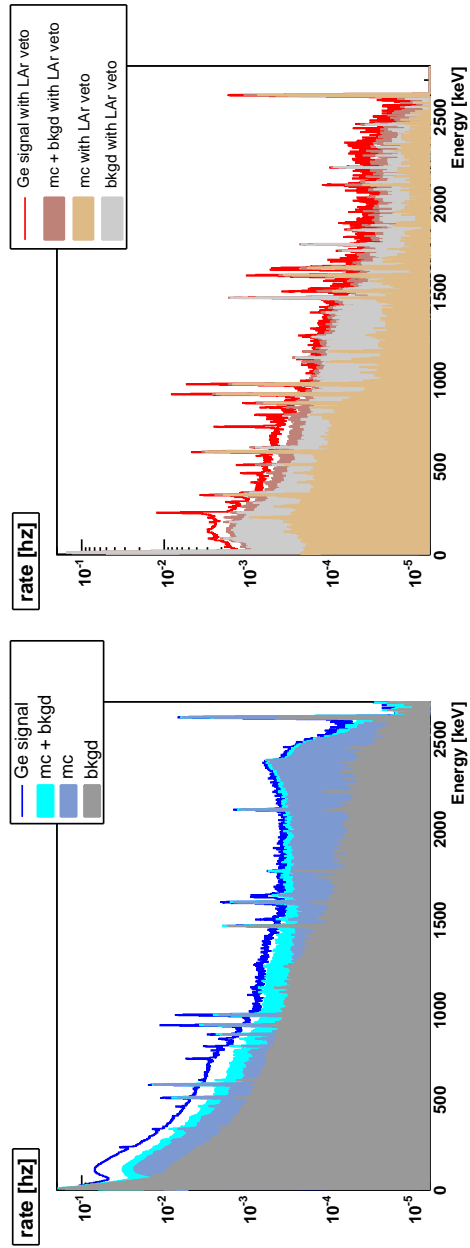}
\caption{Real data compared to \mage\ results: before (left) and after
  (right) background suppression. The source was not in secular equilibrium
thus the poor agreement of MC with data at low energies.}
\label{fig:realmc}
\end{figure}

The experimental data are compared with Monte-Carlo simulations using
the \mage\ Geant4 framework \cite{mage} developed jointly by the
\majorana\ and \gerda\ collaborations.  The spectral shape and
peak-to-Compton ratio measured with $^{137}$Cs, \thorium\ and \ra\ are
well reproduced within typically 5\% or
better. Fig.~\ref{fig:realmc} shows the simulated \thorium\
spectrum. The background spectrum (gray) measured without source is
added to the MC spectrum describing the source.  Note that the
\thorium\ source is not in secular equilibrium, thus the poor
agreement at low energies.
The survival probability after applying the LAr veto cut at 2039~keV
obtained with \mage\ MC simulation corresponds to $S_f = 0.97 $. 
A possible origin of the higher MC value may be related to the
simplified geometrical description of the crystal holder which 
had small LAr dead volumes not included in the simulation.

\section{Conclusion and outlook}
%%%%%%%%%%%%%%%%%%%%%%%%%%%%%%%%

The experimental data show that the detection of LAr scintillation
photons is a powerful method to suppress backgrounds with negligible
loss of \onubb\ signals.  In the setup with an active LAr
mass of 19~kg we observed a background suppression of the \thallium\
Compton continuum at 2039~keV by a factor of 17.  The suppression
factor is limited by gammas escaping from the small LAr 
volume.  
\mage\ MC simulations reproduce the energy spectra as
well as the suppression factors.
An ultra-low background prototype setup is presently under
construction at {\sc Lngs} ({\sc Gerda-LArGe}). The purpose of the
device is to study the novel suppression method at ultra-low
backgrounds with an active LAr mass of approximately 1~ton.
The instrument will be used to study the background of \gerda\ phase~I
detector assemblies prior to their operation in \gerda . Applications 
of the method as an anti-Compton spectrometer for trace analysis 
is envisioned.

\end{document}